\documentclass[conference]{IEEEtran}
\pdfoutput=1
\usepackage{graphicx}
\usepackage{color}
\usepackage{chapterbib}
\usepackage{pdfsync}
\usepackage[hang]{subfigure}

\usepackage{times}
\usepackage{latexsym}
\usepackage{bbm,dsfont}
\usepackage{amsmath}
\usepackage{amssymb}
\usepackage{latexsym}
\usepackage{amsthm}
\usepackage{amsfonts}
\usepackage{epsfig}
\usepackage{latexsym}
\usepackage{cite}
\usepackage{graphicx}
\usepackage{amsmath}
\usepackage{amsthm}
\renewcommand{\figurename}{Fig.}

\newcommand{\bea}{\begin{eqnarray}}
\newcommand{\eea}{\end{eqnarray}}

\usepackage{epsf}
\usepackage{amssymb}
\usepackage{graphicx}
\usepackage{amsmath}
\usepackage[english]{babel}
\usepackage{mathrsfs}
\usepackage{setspace}

\newcommand{\bfl}{\begin{flushleft}}
\newcommand{\efl}{\end{flushleft}}
\newcommand{\bfr}{\begin{flushright}}
\newcommand{\efr}{\end{flushright}}
\newcommand{\bc}{\begin{center}}
\newcommand{\ec}{\end{center}}

\newcommand{\edar}{\end{array}}
\newcommand{\begar}{\begin{array}}

\newcommand{\bit}{\begin{itemize}}
\newcommand{\eit}{\end{itemize}}
\newcommand{\beq}{\begin{equation}}
\newcommand{\eeq}{\end{equation}}
\newcommand{\ben}{\begin{enumerate}}
\newcommand{\een}{\end{enumerate}}
\newcommand{\bean}{\begin{eqnarray*}}
\newcommand{\eean}{\end{eqnarray*}}

\usepackage{ifthen}
\usepackage{color}
\usepackage{amssymb}

\usepackage[newenum]{paralist}

\newcommand{\IsTR}{\ifthenelse{1<2}}

\IsTR{}{ \ninept }

\usepackage{wrapfig}
\usepackage{algorithm}
\usepackage{algorithmic}
\usepackage{stfloats}
\newcommand{\green} {\color{black}}

\begin{document}

\title{Distributed High Accuracy Peer-to-Peer Localization in Mobile Multipath Environments}

\author{ \IEEEauthorblockN{Venkatesan. N. Ekambaram and Kannan Ramchandran}
\IEEEauthorblockA{Department of EECS, University of California, Berkeley\\
Email: \{venkyne, kannanr\}@eecs.berkeley.edu}}

% make the title area
\maketitle

\begin{abstract}
In this paper we consider the problem of high accuracy localization of mobile nodes in a multipath-rich environment where sub-meter accuracies are required. We employ a peer to peer framework where the vehicles/nodes can get pairwise multipath-degraded ranging estimates in local neighborhoods together with a fixed number of anchor nodes. The challenge is to overcome the multipath-barrier with redundancy in order to provide the desired accuracies especially under severe multipath conditions when the fraction of received signals corrupted by multipath is dominating.  We invoke a message passing analytical framework based on \emph{particle filtering} and reveal its high accuracy localization promise through simulations.\\

\emph{Key words-} Localization, Multipath, Graphical Models, Particle Filtering, Hidden Markov Model.
\end{abstract}

\IEEEpeerreviewmaketitle

\section{Introduction}
High-accuracy localization is mandated in many applications like vehicle safety  \cite{wilson1998potential},  autonomous robotic systems \cite{nerurkar2009distributed},   Unmanned Air Vehicle (UAV) systems etc, where sub-meter accuracies are called for. Standard GPS receivers can have errors over fifty or more meters which is unacceptable for many of these applications. The principal problem is multipath \footnote{Multiple delayed versions of the {\green same} transmitted signal {\green are} received at the receiver due to reflections from different objects in the environment.} interference \cite{chen1999non}, which is particularly prevalent in cities and ``urban canyon'' environments.  In a multipath-rich environment, the received signals are no longer gaussian in nature challenging the use of standard estimation techniques like the well-known Kalman Filtering framework and its extensions. The principal idea behind GPS is to obtain three or more distance measurements from sources with known locations (e.g. satellites) and estimate the location based on trilateration. However, even if one of the measurements is corrupted by multipath, the location errors can be significantly large. It has been well noted that redundancy in measurements is the key to tackle multipath \cite{chen1999non}. Many of the existing solutions such as D-GPS, A-GPS \cite{djuknic2001geolocation} augment the GPS system by adding extra infrastructure in terms of fixed base stations. However infrastructure cost and complexity constrains the amount of redundancy that can be introduced in the system and as a consequence limits the achievable localization.

\begin{figure}
\centering
\vspace*{-0.25in}
\includegraphics[height=1.5in]{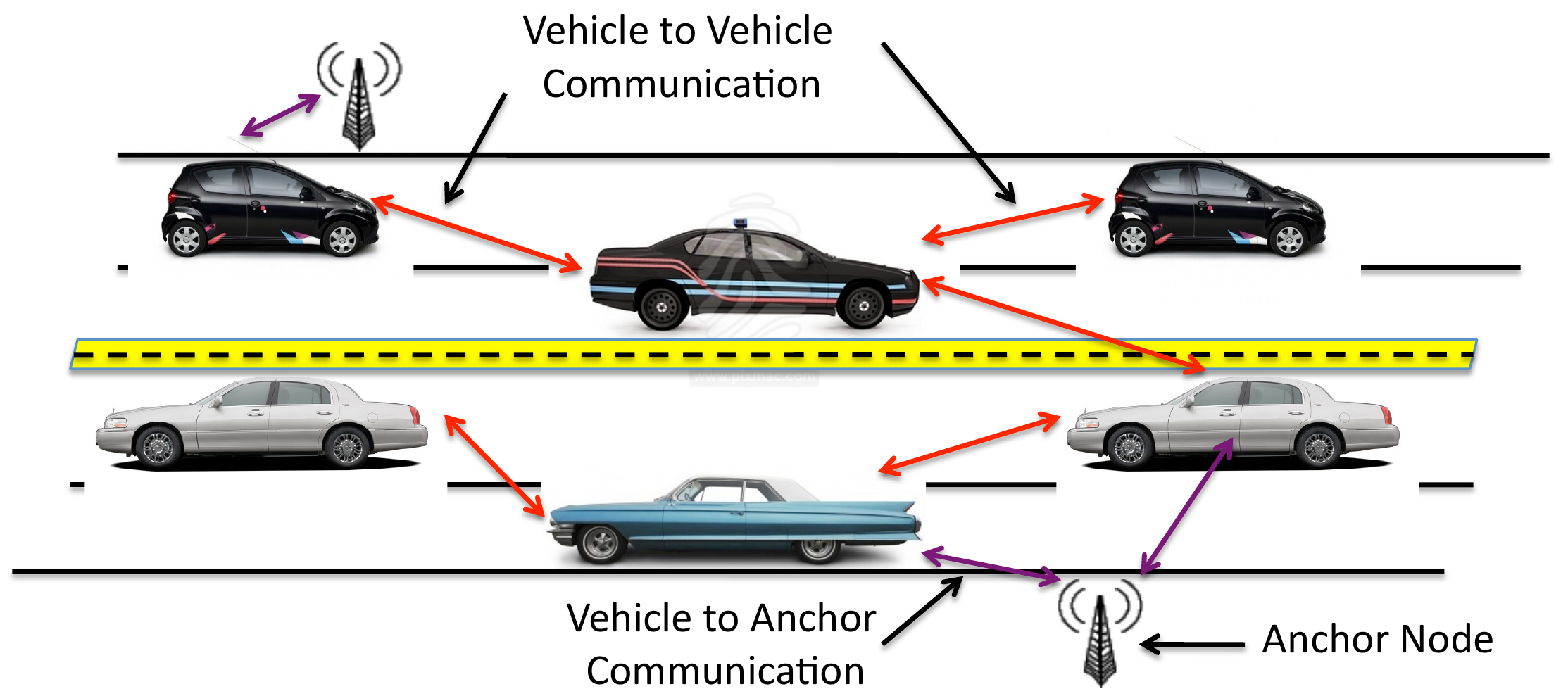}
\caption{ Peer-to-Peer Collaborative Localization}
\label{fig:vanet}
\vspace*{-0.5in}
\end{figure}

In contrast to existing centralized systems that are based on a ``cellular-like'' architecture, with users individually computing their location by  calculating the distance to a small number of satellites and/or terrestrial base stations, we adopt a ``peer to peer'' architecture (see Fig \ref{fig:vanet}) where nodes collaborate and help each other to refine their position estimates. We look for distributed algorithms in the interest of scalability and reduced computational complexity. We propose the inclusion of low-cost static ``anchor'' nodes with known locations and study the effect of the number of anchor nodes and existing mobile nodes on the system performance. Collaboration coupled with mobility generates a large pool of measurements in the system. The fundamental insight is that, some fraction of these measurements will be produced by line-of-sight (LOS) dominated signals, and hence be fairly accurate, while some fraction will be corrupted by dominated non-line-of-sight (NLOS) reflected waves.  Receivers do not know a priori which measurements are LOS and which are NLOS. Hence, the task of the users is to cooperatively discard the NLOS signals, thus enabling them to compute high-precision position estimates. Our main contribution in this paper is to uncover a framework and a distributed algorithm founded on \emph{message passing} for collaborative localization. We adopt a mixture model (discussed in Sec. \ref{sec:ProbSet}) for the received signal to characterize LOS/NLOS, that naturally arises given that the receivers do not know a priori the nature of the received signal.

Distributed algorithms \cite{nerurkar2009distributed}, \cite{patwari2005locating} have been of recent interest for collaborative localization. However, most of the existing work in the literature on localization focus on the case when the measurements are gaussian. In our setting, the distributions of interest are mixture distributions that are highly non-gaussian in nature. Generalizations of Kalman Filtering, such as the Extended Kalman Filter (EKF) \cite{kiriy2002three} have been proposed for non-gaussian problems. However, EKF solutions do not work well when the distributions are bimodal as in the case of mixture distributions,  since gaussians do not approximate bimodal distributions very well. Some of the other approaches try to weed out the NLOS signals  (\cite{chen1999non},  \cite{ananthasubramaniam2008cooperative}, \cite{casas2006robust}, \cite{seow2008non}), and work only with the LOS signals which can be reasonably modeled as gaussian. The underlying algorithm used in most of these approaches has been modified forms of the Random Sample Consensus (RANSAC) algorithm, which is a classical algorithm in computer vision literature to discard outliers in the data. RANSAC works well when the number of outliers (here the NLOS signals) is much smaller than the number of LOS signals.  However, these are limited in settings (for e.g. see  \cite{renzo2006ultra} for an ultra-wideband setting) where the fraction of LOS signals is typically lesser than NLOS which would challenge the performance of RANSAC-based approaches. The main reason for this is that in our case  the NLOS signals dominate the set of measurements, and those can no longer be considered as outliers. Simulation results show that our algorithm works very well even in this challenging case where the fraction of LOS signals is dominated by the fraction of NLOS signals. An excellent survey of existing locationing algorithms and their drawbacks can be found in \cite{boukerche2008vehicular}.

%Hack to have this equation at the end of next page.
\begin{figure*}[b]
\vspace*{-0.25in}
\begin{eqnarray}
\hline
\hspace*{-0.25in}
p(\{v_t\}, \{\theta_t\}, \{z_t\} )  =  \prod_{t,k,m}p(\theta_t(km)|v_t(k),v_t(m),z_t(km)) \prod_{k=1}^N p(v_1(k))\prod_t p(v_t(k)|v_{t-1}(k)  \prod_{k,m} p(z_1(km))\prod_t p(z_t(km)|z_{t-1}(km)).                                               
\end{eqnarray}
\end{figure*}

\section{Problem Setup}
\label{sec:ProbSet}
We have $N$ mobile nodes with an arbitrary mobility model and $M$ static anchor nodes with known locations. For simulations we consider the nodes to be moving with a constant velocity along a fixed trajectory, motivated by the highway setting for vehicles. Each vehicle is equipped with a sensor capable of getting a Time of Arrival (ToA)/ Time Difference of Arrival (TDoA) signal from other vehicles/anchors within a communication radius $R$. {\green We model each measurement as  either a LOS-dominated signal or an NLOS-dominated signal by choosing the observation noise in the received signal to be drawn from a mixture of two distributions, corresponding to LOS and NLOS respectively.}  The model is motivated by some of the experimental work  carried out in the UWB  \cite{renzo2006ultra}, \cite{turin1972statistical}, \cite{pedersen2000stochastic}. For e.g., the experimental results in \cite{renzo2006ultra} show that some fraction of the received signals are purely LOS-dominated signals which motivates the mixture distribution. For simulations, we will model the noise in LOS as gaussian whereas for the NLOS we will take it to be a sum of an exponential and a gaussian distribution. The noise model for NLOS is again motivated from some of the experimental results discussed in \cite{turin1972statistical}, \cite{pedersen2000stochastic}. These show that the NLOS signal distribution is very close to an exponential distribution. This was earlier conjectured in \cite{turin1956communication}, where the author argues that the arrivals of the different paths can be modeled as a Poisson process which in turn leads to the inter arrival times being exponential.  The theory developed here is however, more generic and does not depend on the specific nature of the distributions. Each vehicle is assumed to be equipped with an accelerometer and magnetometer  that give noisy readings of the velocity and direction of motion of the vehicle. We will assume that there is a multiple access protocol in place that will help the vehicles communicate across the shared medium. 

  Let $v_t(k)$ denote the true  location of the $k$th vehicle at time instant $t$. $\theta_t(km)$ denotes the reading (distance measurement) between the vehicle $k$ and node\footnote{The term vehicle/anchor and node will be interchangeably used.} $m$ at time instant $t$.  The readings $\theta_t(km)$ are sampled  from a mixture of two distributions, ($p_{LOS}(\theta_t(km)|v_t(k),v_t(m)),p_{NLOS}(\theta_t(km)|v_t(k),v_t(m)))$),
with mixture probabilities $(\alpha,1-\alpha)$ respectively.   Let $z_t(km)$ be the indicator random variable for the LOS reading between vehicles $k$ and $m$ at time instant $t$.
 \bean
  z_t(km) & = & \left\{
\begin{array}{rl} 1 & \mbox{if }\theta_t(km) \sim p_{LOS}(\theta_t(km)|v_t(k),v_t(m)) \\
                              0 & \mbox{otherwise}\
 \end{array} \right. 
 \eean
  It is highly unlikely that a vehicle's state would change rapidly over time. To capture this, we will model the state evolutions of the  $z_t(km)$'s as a ``sticky'' Markov Chain with a stationary distribution  $(\alpha, 1-\alpha)$. Let $p(v_t(k)|v_{t-1}(k))$ be the distribution that governs the evolution of the vehicle states across time which is an artifact of the inertial navigation system.
  
 Every vehicle $k$ needs  to estimate its location  $v_t(k)$, based on all measurements $\{\theta_{\tau}(km)\}_{\tau = 1}^t$ from its neighbors upto time $t$.  Given the non-gaussian nature of the problem and the accuracy requirement we need to develop robust, accurate and distributed algorithms.   \emph{Particle Filters} have gained importance in the recent past for tackling non-gaussian estimation problems. These provide accuracies close to Minimum Mean Square Error (MMSE) estimates. The nature of our problem helps us obtain Kalman-like updates for particle filtering giving rise to a simplified algorithm. A short primer on graphical models and particle filtering is provided in the next section.
  
\section{Graphical models \& Particle Filtering Primer}
Graphical models and particle filtering have been extensively studied in the machine learning community. An excellent treatment of graphical models is provided in  \cite{jordan1998learning} and discussions on particle filtering can be found in \cite{doucet2001sequential}. We describe these briefly to make our paper  self contained.
\subsection{Graphical models}
\begin{figure}
\vspace*{-0.5in}
\centering
\includegraphics[height=2.5in]{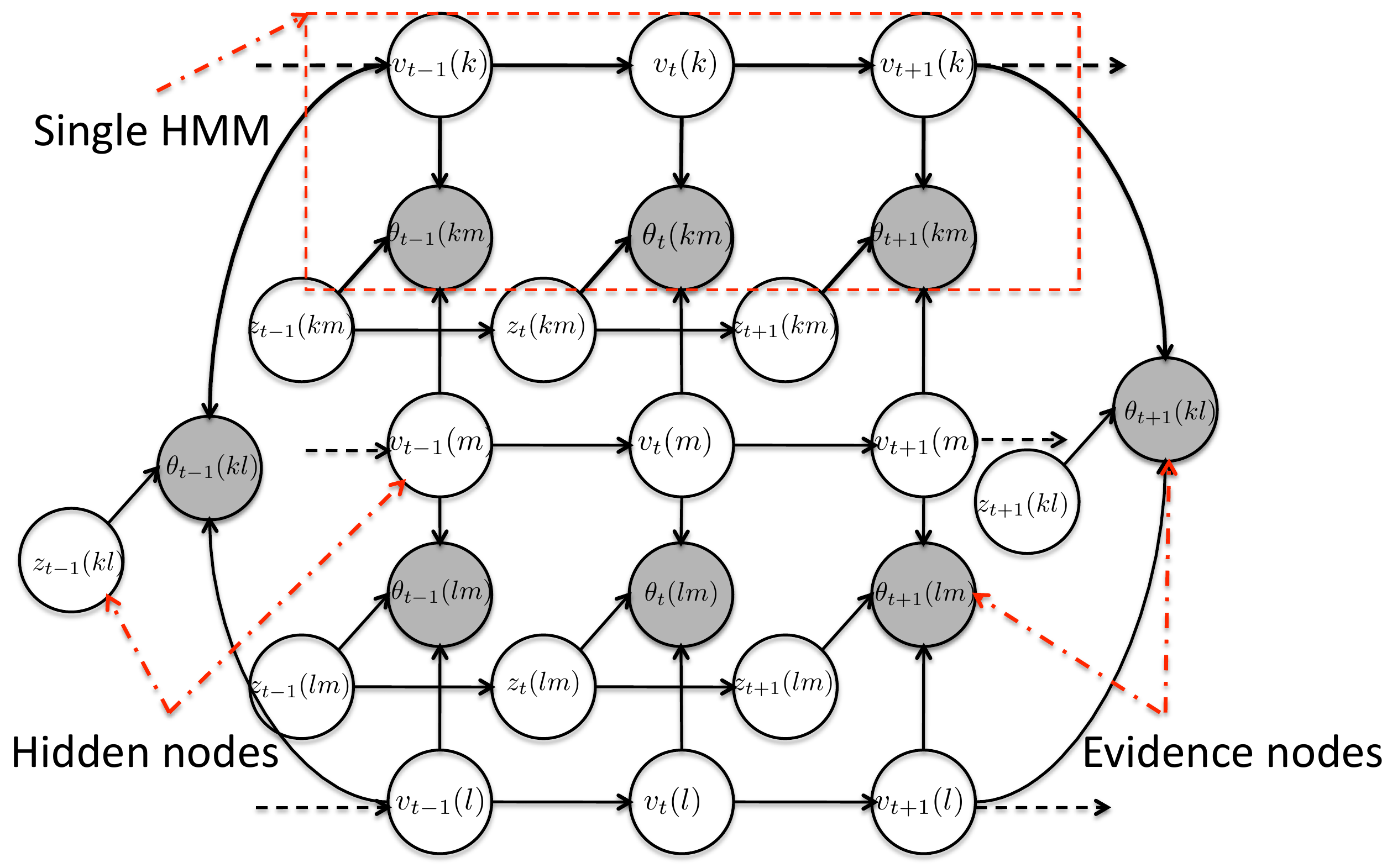}
\caption{Graphical model representation of the unknown vehicle states (locations) and the observations (LOS/NLOS measurements).}
\label{fig:VehHmm}
\vspace*{-0.4in}
\end{figure}
Graphical models provide a good way of understanding dependencies between different random variables. These models come in handy when we are trying to estimate some parameters given probability 
distributions that have a certain structure and we need computationally feasible algorithms to get the estimates. A directed acyclic graph, $\cal{G}(V,E)$, consists of a vertex set ${\cal V}$ and edge set ${\cal E}$ which is the collection of all directed edges. The vertices (nodes) of the directed graph represent random variables and the {\green dependencies} amongst the random variables are captured by the edges. Let $\{X_s, s \in {\cal V}\}$ denote the set of all random variables indexed by the nodes of the graph. For every node $s \in {\cal V}$, let $\pi_s$ denote the set of indices of its parents \footnote{{\green Node $p$ is the parent of node $s$, if there is an incoming edge to $s$ from $p$.}}. {\green For any $S \subseteq {\cal V}$, let $X_S \triangleq \{X_s, s \in S\}$}. Then, the joint probability distribution of the random variables can be factored as 
$p(X_{\cal V})  =    \prod_{s \in {\cal V}} p(X_s | X_{\pi_s}).$

Dependencies amongst the vehicle locations and the readings are captured by the graphical model shown in Fig \ref{fig:VehHmm}, which is a coupled Hidden Markov Model (HMM). The unshaded nodes are the \emph{hidden nodes}  to be estimated \footnote{Anchor nodes are not shown in this model for simplicity.} and the shaded nodes are observations coupling the different Markov chains of the vehicles, called as the \emph{evidence nodes}.  Based on the factorization described above, the joint probability distribution of the set of all random variables is given by equation (1).

The primary goal here is to estimate the hidden nodes given the observations. It is hard to directly apply conventional inference algorithms such as the celebrated loopy belief propagation \cite{jordan1998learning} considering the fact that the hidden states are continuous in nature and that the graph grows over time.  We will introduce approximations to the graphical model and use particle filtering to do inference over the approximated model. Simulation results show that we can obtain high accuracies in spite of the approximations. We briefly describe particle filters in the next section.

\subsection{Particle Filtering}
 \begin{figure}
 \vspace*{-0.4in}
 \centering
 \includegraphics[height=1.25in]{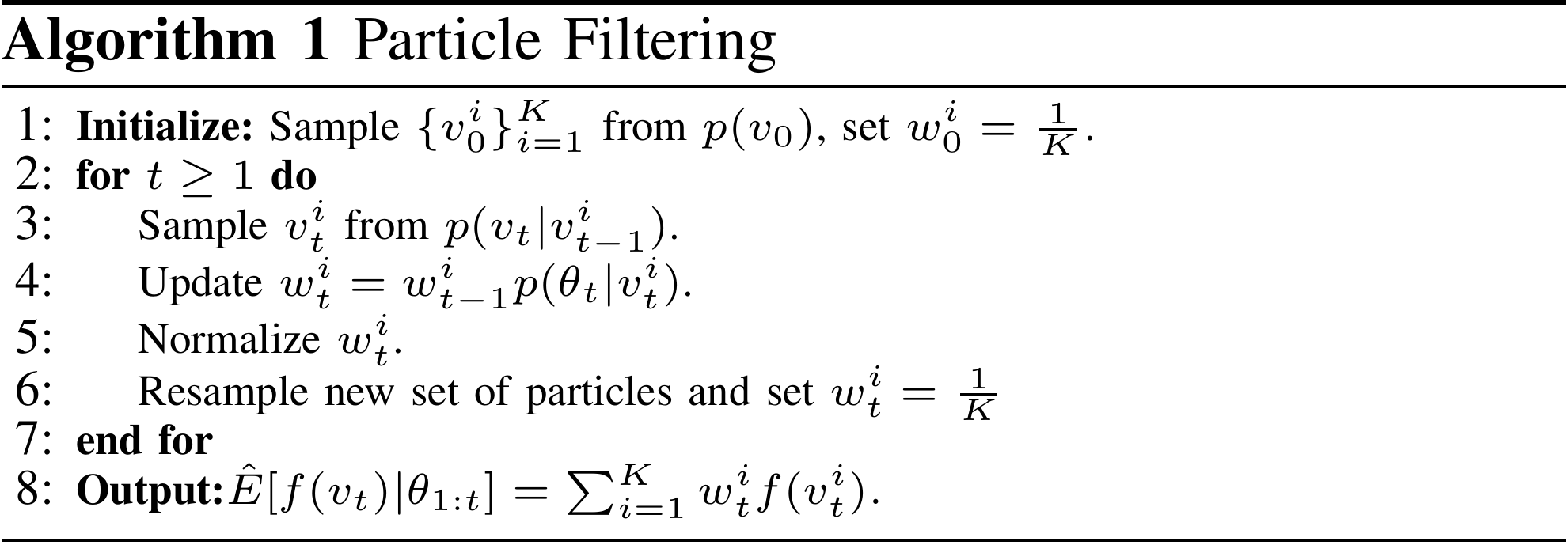}
 \vspace*{-0.75in}
 \end{figure}
Particle Filtering is a Monte Carlo simulation technique whose goal is to approximate
the posterior state density that can be used to obtain the MMSE estimate. We will consider only a single vehicle for now and omit the vehicle index $k$ for notational simplicity. Let $v_{1:t}$ and $\theta_{1:t}$ denote
the set of states and observations up to time $t$ respectively. The posterior density of the state given the observations can be approximated as 
$p(v_t|\theta_{1:t}) \approx \hat{p}(v_t|\theta_{1:t}) = \frac{1}{K}\sum_{i=1}^K \delta(v_t - v_t^{i})$
where $\{v_{t}^{i} ; i=1,...,K\}$ are K i.i.d random samples (particles) picked from the distribution $p(v_t|\theta_{1:t})$ and $\delta(.)$ is the dirac-delta function. Expectations of some functions of the state $f(v_t)$, can then be approximated with high probability, for large $K$, as 
$ E[f(v_t)|\theta_{1:t}] \approx \frac{1}{K}\sum_{i=1}^K f(v_t^i).$
\begin{figure}
\vspace*{-1in}
\centering
\includegraphics[height=2.9in]{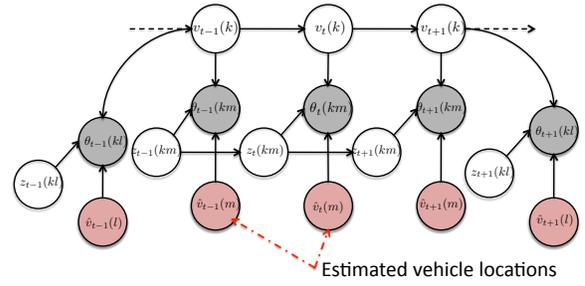}
\vspace*{-1in}
\caption{Approximation for the original coupled HMM.}
\label{fig:VehHmmApp}
\end{figure}
Typically sampling from the {\green true} posterior distribution is hard. {\green Let us consider sampling from a distribution  $\pi(v_{1:t}|\theta_{1:t})$, known as   \emph{proposal distribution}}. After some simple manipulations using Bayes rule, we get
\bean
 E[f(v_{1:t})|\theta_{1:t}] & = & \frac{E_{\pi}[w_t(v_{1:t}) f(v_{1:t})]}{E_{\pi}[w_t(v_{1:t})]},\\
 \eean
where $w_t(v_{1:t})$ are known as the importance weights given by
\[w_t(v_{1:t}) = \frac{p(\theta_{1:t}|v_{1:t})  p(v_{1:t})}{\pi(v_{1:t}|\theta_{1:t})}.\]
Thus, if we have i.i.d samples of $v_{1:t}$ from the proposal distribution  $\pi(v_{1:t}|\theta_{1:t})$, we get
$
E[f(v_{1:t})|\theta_{1:t}] \approx \sum_{i=1}^K \tilde{w_t}^i f(v_{1:t}^i),
$
 where $\tilde{w_t}^i$ are the normalized importance weights. {\green Convergence of this estimate to the true MMSE estimate is shown in \cite{doucet2001sequential}}.
 
  For the case of a single HMM (Fig \ref{fig:VehHmm}), we have
  $ p(v_{1:t}, \theta_{1:t}) = p(v_1) \prod_{j=2}^t p(v_j|v_{j-1})  \prod_{j=1}^t p(\theta_j|v_j).$
Lets choose a proposal distribution that admits a decomposition
$\pi(v_{1:t}|\theta_{1:t}) = \pi(v_{1:t-1}|\theta_{1:t-1}) \pi(v_t|v_{1:t-1} \theta_{1:t}).$
After simplifications, we can obtain a recursive estimation for the weights,
\[ w_t = \frac{w_{t-1}p(\theta_t|v_t)p(v_t|v_{t-1})}{\pi(v_t|v_{1:t-1} \theta_{1:t})}.\]  The optimal proposal distribution, that minimizes the variance of the error has been shown \cite{doucet2001sequential} to be $p(v_t|v_{t-1},\theta_t)$. Since sampling from this distribution is difficult for our problem, we will use a simpler distribution {\green $ \pi(v_t|v_{1:t-1} \theta_{1:t}) = p(v_t|v_{t-1})$} to get the weight updations $w_t = w_{t-1}p(\theta_t|v_t)$.  The reader is referred to the literature for other methods of choosing a proposal distribution. Typically a resampling step is introduced (see Alg. 1), to handle degeneracy issues when particle weights become too low.

\section{Particle Filtering for Localization}

Exact inference over  Fig \ref{fig:VehHmm} being hard,  we will resort to an approximation for every vehicle as shown in Fig \ref{fig:VehHmmApp}. The new set of shaded nodes in this graph correspond to the estimated location of the other vehicles. At every time instant each vehicle gets the estimated location of its neighbors from the previous time instants and assuming that it is close enough to the true location, the vehicle gets an estimate of its own location using particle filtering. A straightforward method of particle filtering over this model would be to consider $\{v_t,z_t\}$ as a random variable pair which would reduce the graphical model to a simple HMM. However the state space of the random variables grows exponentially with the number of neighbors as we now need to sample particles over $\{v_t(k),z_t(km)\}$, which could lead to scaling issues. More importantly particle filtering is efficient when the hidden states are continuous, whereas $z_t$'s are binary. Thus we now see to combine particle filtering and exact inference to obtain simplified weight updations.

Consider the HMM in Fig \ref{fig:VehHmmApp} and ignore the vehicle indices $k$ and $m$.
Using the same definition of $w_t(v_{1:t})$ as before, we can now write $w_t(v_{1:t}) = \sum_{z_t} \phi(v_{1:t},z_t)$, where 
\[\phi(v_{1:t},z_t)  =  \frac{p(\theta_{1:t}, z_t|v_{1:t})  p(v_{1:t})}{\pi(v_{1:t}|\theta_{1:t})}.\]
Assuming a similar proposal distribution factorization as before we can show that 
\[
\hspace*{-0.1in}
\phi(v_{1:t},z_t)  =  \frac{p(\theta_{t}|z_t,v_t) p(v_t | v_{t-1})}{\pi(v_t|v_{1:t-1} \theta_{1:t})} 
                             \sum_{z_{t-1}} p(z_t|z_{t-1}) \phi(v_{1:t-1},z_{t-1}).
\]
Choosing $\pi(v_t|v_{1:t-1} \theta_{1:t}) = p(v_t | v_{t-1})$, we get 
\bean\phi_t \triangleq \phi(v_{1:t},z_t)  = p(\theta_{t}|z_t,v_t) \sum_{z_{t-1}} p(z_t|z_{t-1}) \phi(v_{1:t-1},z_{t-1}).\eean
For each vehicle $k$ and its neighbor $m$ we have $\phi_t(km)$ and the update equations are given in the Algorithm 2.

 The optimal detection rule for $z_t$ is given by 
 \bean
  \hat{z}_t & = & \left\{
\begin{array}{rl} 1 & \mbox{if } p(z_t = 1| \theta_{1:t}) >  p(z_t = 0| \theta_{1:t})\\
                              0 & \mbox{otherwise}\
 \end{array} \right. . 
 \eean 
 This can be simplified and shown to be equal to the following test 
$E_{\pi} \phi(v_{1:t},z_t=1)  \gtrless^{1}_{0}  E_{\pi} \phi(v_{1:t},z_t=0)$,
which is evaluated by approximating $E_{\pi} \phi(v_{1:t},z_t)  \approx \sum_i \phi(v_{1:t}^i,z_t)$.  
   \begin{figure}
 \vspace*{-0.5in}
 \centering
 \includegraphics[height=2in]{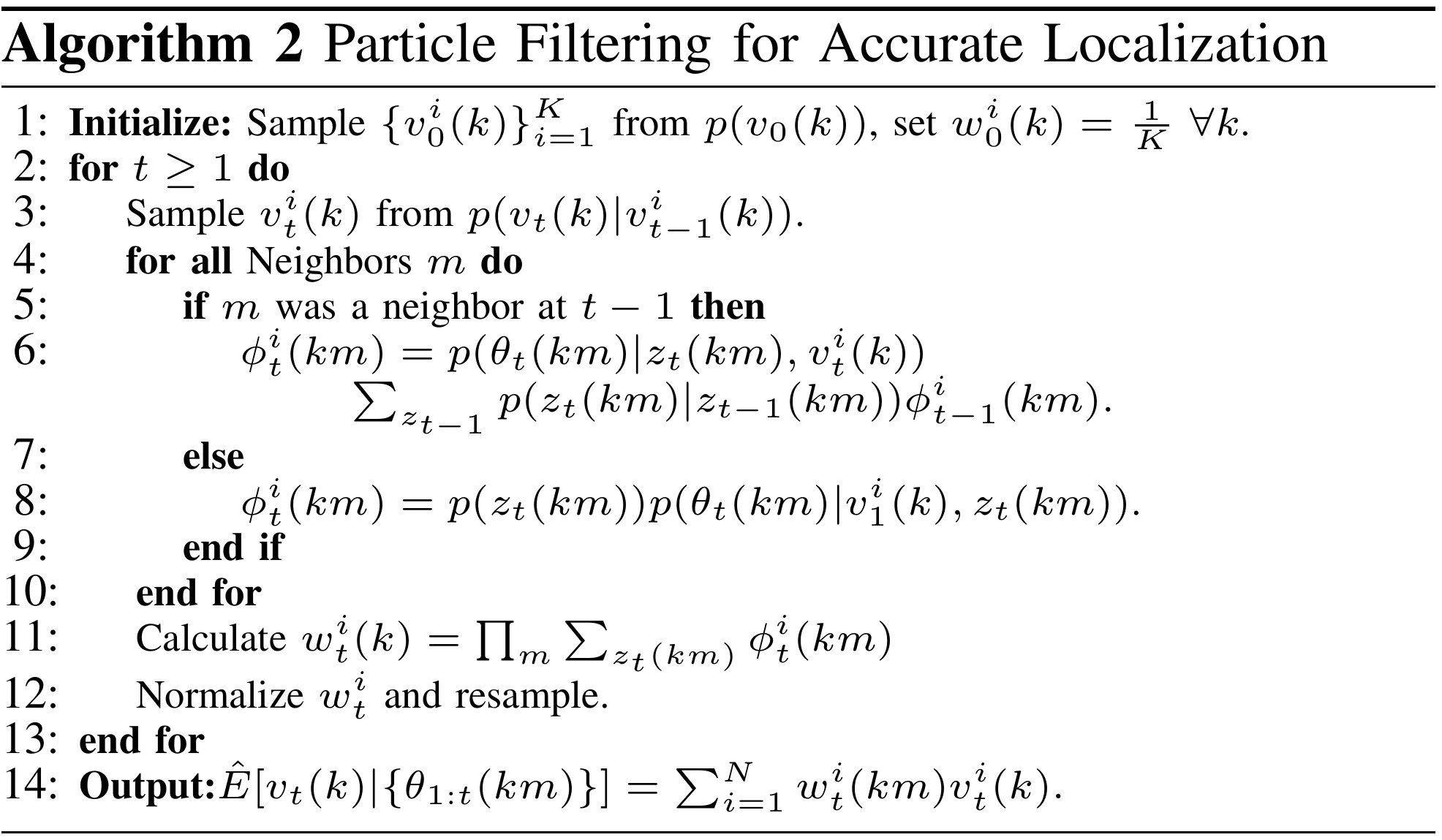}
 \vspace*{-0.6in}
 \end{figure}
To take care of degeneracy issues over long time instants \cite{doucet2001sequential}, we reset the system when the number of distinct particles become small and sample a new set of particles from a neighborhood around  the location estimate at that time instant.
 \section{Simulation results}
% 
% \begin{table}[htdp]
%\begin{center}
%\begin{tabular}{|c|c|c|}
%\hline
%Num of Anchors & Num of Vehicles & Mean Error\\
%(M) & (N) & (Particle Filtering)\\
%\hline
%26                       &  20     & 0.93 m\\
%26                     &   24   & 0.82 m\\
%26                    &    30    &0.76 m\\
%30                    &   20         &0.74 m \\
%36                    & 20 & 0.70 m\\
%\hline
%\end{tabular}
%\end{center}
%\caption{$P_d$ of the hidden $z$ states and the mean estimation error.}
%\label{tab:PdLos}
%\vspace*{-0.3in}
%\end{table}
%\begin{table}[htdp]
%\begin{center}
%\begin{tabular}{|c|c|c|c|}
%\hline
%$\alpha$ ($\%$) & $P_d$ & Mean Error & Mean Error \\
%                              &               &  (Particle Filtering) & (Maximums Likelihood)\\
%\hline
%5                       &  0.43     & 1.04 m & 1.05 m\\
%15                     &   0.52   & 0.99 m&1.01 m\\
%30                    &      0.61    & 0.93 m& 0.94m\\
%45                    &        0.68         &0.71 m &  0.73 m \\
%\hline
%\end{tabular}
%\end{center}
%\caption{Mean estimation errors of the proposed particle filtering algorithm and maximum likelihood as  a function of the fraction of LOS signals ($\alpha$). Also shown is the detection probability of the LOS signals ($P_d$).}
%\label{tab:PdLos}
%\vspace*{-0.3in}
%\end{table}
   \begin{figure}[h]
\centering
\vspace*{-0.6in}
%\hspace*{-0.3in}
\subfigure[Cumulative density function of localization error as a function of $\alpha$.]
{\includegraphics[height =2.25in, width = 3.5in]{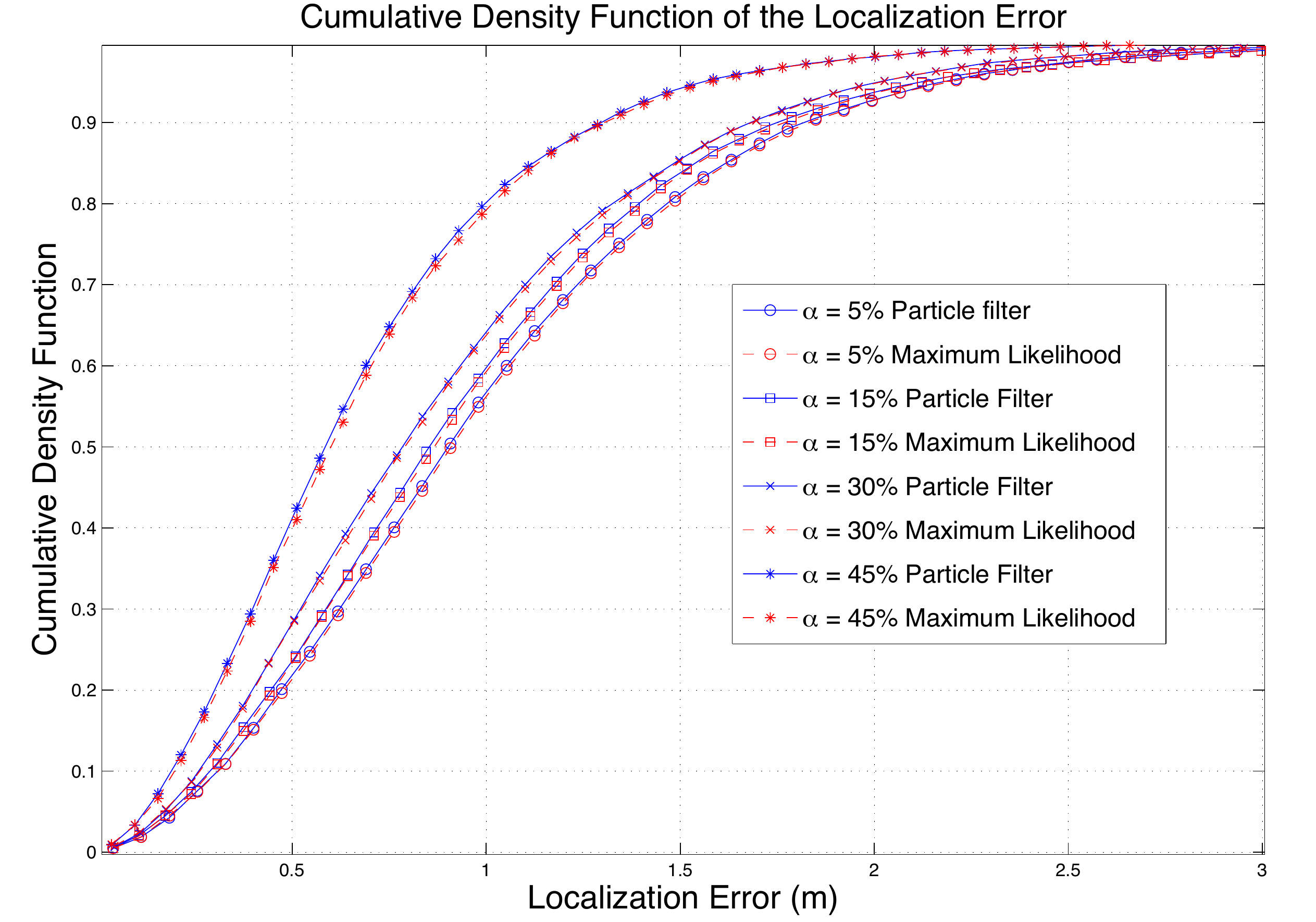} \label{fig:ErrorCdf}}\\
%\hspace*{-0.3in}
\subfigure[True and estimated vehicle trajectories for the proposed algorithm, $\alpha = 15\%$]{\includegraphics[height =2.5in]{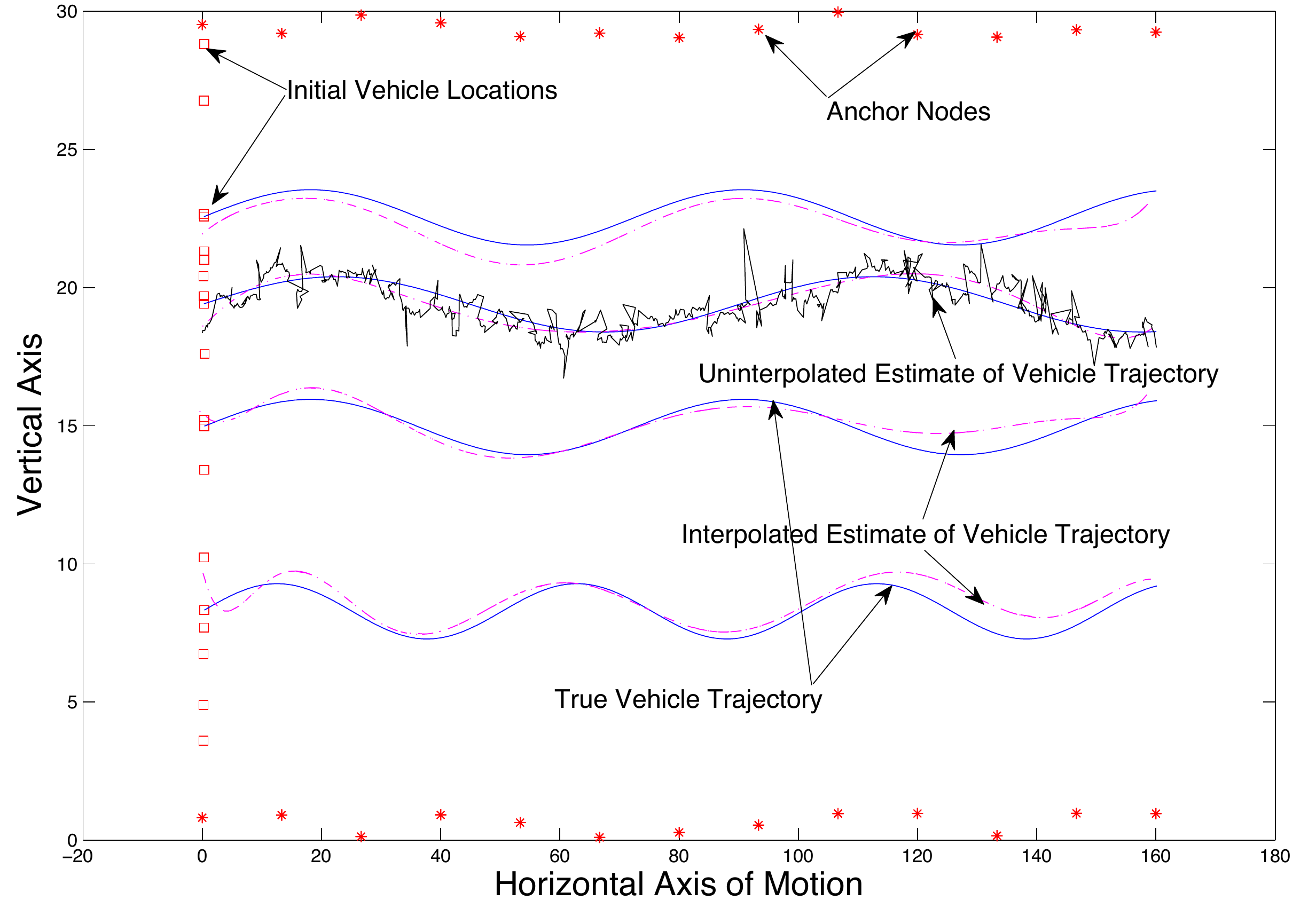}
\label{fig:TrajVeh}}\\
\subfigure[Mean estimation errors and the detection probability of the LOS signals ($P_d$) as  a function of $\alpha$. ]{\includegraphics[height =1in]{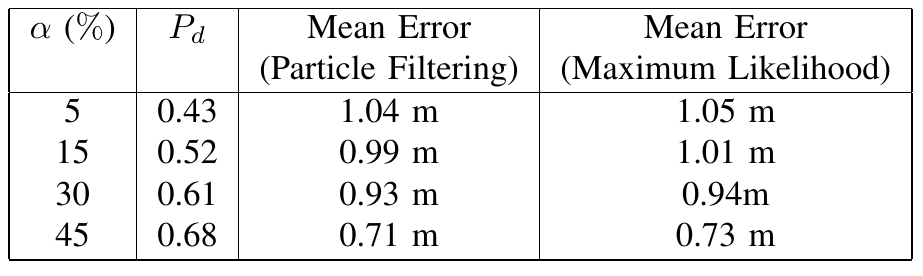}
\label{tab:ErrorTable}}\\
\subfigure[Mean estimation errors of the particle filtering algorithm as a function of the number of anchor nodes and vehicles for  $\alpha = 30\%$.]{\includegraphics[width = 3.5in, height=1in]{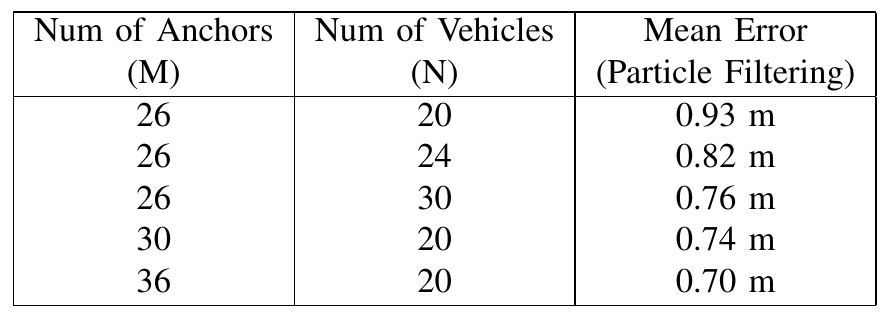}
\label{tab:NumAncVeh}}\\
\caption{Parameters for all the plots: $\sigma_{INS} = 0.1$, $\sigma_{LOS} = 0.05$, $\sigma_{NLOS} = 5$, Num of particles $= 900$, Horizontal velocity$ = 0.2m$ per time step $\forall k$, $M = 26$, $N = 20$, $R = 10m$.  The plots are a function of the fraction of LOS signals $\alpha$. The proposed particle filtering algorithm is compared against the local Maximum Likelihood algorithm.}
\vspace*{-0.3in}
\end{figure}

 The simulation set up consists of $N = 20$ vehicles moving in a grid of size $150m \times 30m$. The vehicles start at random locations from the left most corner of the grid and move at a constant velocity to the right. To account for the vertical motion, a simplistic model of a curved trajectory is simulated for each vehicle. The observations are generated as follows. If $d_t(km) = ||v_t(k)-v_t(m)|| < R$,
  \bean
  \theta_t(km) & = & \left\{
\begin{array}{rl} d_t(km)+ n_t(km) & \mbox{if  LOS}\\
                              d_t(km) + \epsilon_t(km) + n_t(km)  & \mbox{otherwise}\
 \end{array} \right. 
 \eean 
 where $n_t(km) \sim N(0,\sigma_{LOS}^2)$ i.i.d, $\epsilon_t(km) \sim Exp(\sigma_{NLOS}^{-1})$.  We assume the fraction of the readings that are LOS to be $\alpha$. The evolution of the $z_t(km)$ random variables is governed by the probability law,
 $p(z_t(km) = 1|z_{t-1}(km)=0) = \frac{\alpha}{2}$ for these simulations and the other values are taken so that the stationary distribution is $(\alpha, 1-\alpha)$. The inertial navigation system readings are assumed to be obtained under an additive white gaussian noise model. Time steps are divided into units of one for simplicity. We compare the performance of the algorithm to a local genie aided Maximum Likelihood (ML)  algorithm. Here each vehicle, at every time instant, calculates the local ML estimate of its location assuming that a genie provides it with the exact locations of its neighbors.
 
 The localization error cumulative density function is plotted for different values of $\alpha$ in Fig \ref{fig:ErrorCdf}. The $x$-axis is the set of error values and $y$-axis is the cumulative density function. One can see that even at low values of $\alpha = 5\%$, more than $80\%$ of the errors are less than $1.5m$. The algorithm performance is slightly better than ML algorithm. This is not too surprising considering that particle filtering tries to approximate MMSE which is the optimal solution for mean squared loss function. Algorithms based on RANSAC were found to have errors over $5m$ and are not discussed here. The true and estimated vehicle trajectories are plotted in Fig \ref{fig:TrajVeh}. The estimated vehicle trajectory is generated by a simple polynomial fit to all the estimated vehicle locations. Table \ref{tab:ErrorTable} shows the probability of detection ($P_d$) of the hidden  $z_t(km)$ states (fraction of the times LOS states are detected correctly) and the mean estimation error.  The strength of the algorithm is in exploiting the large pool of measurements efficiently where algorithms like RANSAC fail. Table \ref{tab:NumAncVeh} shows the localization error performance as a function of the number of anchor nodes and mobile nodes in the system. One can observe a law of diminishing returns as the number of anchor nodes increases.  Another interesting observation is that the percentage improvement in the localization performance when the number of mobile nodes, that act like  ``pseudo-anchors'', is the similar to that of adding more anchors.  A theoretical exploration of this phenomenon is currently under progress.
 
We shall now discuss some practical aspects of implementing this scheme. The communication between the nodes can be carried over the Dedicated Short Range Communications (DSRC) band, that has been allocated by the Federal Communications Commission (FCC) for Intelligent Transportation Systems. IEEE 802.11p can be used for medium access.  This is essentially a CSMA protocol specialized for vehicular networks. Whenever a node receives a 802.11p packet from some other node, it can estimate the time delay of arrival that translates to a distance measurement used for localization. The location estimate sent by one node to the other node can be sent as part of the 802.11p packet. The parameters of the distributions and the fraction $\alpha$ can be estimated using an Expectation Maximization (EM) algorithm at every step. Each vehicle has an average of 8 other vehicles and 4 anchor nodes in its communication radius which is quite a reasonable assumption. In practice, the anchors usually have a larger range of communication and the anchor density required would be lesser than that used in the simulations. We assume that the computational load in updating the particles is manageable by present day vehicles that have a reasonably high processing power.
 
 \section{Conclusion and Future work}
In this work we explored the application of particle filtering to get estimates of vehicle locations in a highly NLOS environment. We derived weight update equations for the NLOS setting and simulation results show that reasonably good accuracies in positioning is feasible. Future work includes carrying out more realistic simulations using traffic and network simulators. The approximation in the graphical model could break down above a certain noise threshold and below a certain anchor density, and the algorithm could potentially diverge. A theoretical understanding of when the algorithm diverges is another research direction, though we believe this to be a hard problem. A theoretical exploration of the effect of increased anchor and mobile nodes is under progress. Integrating other sensing modalities into the algorithm is a future research direction.

%\section*{Acknowledgment}
%
%The authors would like to thank the anonymous reviewers and the TPC reviewers for their detailed and useful comments that helped improve the quality of the paper.

\bibliographystyle{IEEEtran}
\bibliography{GBComBib}

%\begin{thebibliography}{1}

%%\bibitem{IEEEhowto:kopka}
%%H.~Kopka and P.~W. Daly, \emph{A Guide to \LaTeX}, 3rd~ed.\hskip 1em plus
%%  0.5em minus 0.4em\relax Harlow, England: Addison-Wesley, 1999.

%\end{thebibliography}

\end{document}